\documentclass[
    ,final            
  ]
  {aipproc}

\layoutstyle{8x11double}

\begin{document}

\title{Heavy mesons in dense matter}


\author{Laura Tolos$^{1,2}$, Daniel Gamermann$^3$, Carmen Garcia-Recio$^4$, Raquel Molina$^5$, \\ Juan Nieves$^5$, Eulogio Oset$^5$,  Angels Ramos$^3$}{
  address={$^1$Theory Group. KVI. University of Groningen,
Zernikelaan 25, 9747 AA Groningen, The Netherlands \\
$^2$ Instituto de Ciencias del Espacio (IEEC/CSIC), Campus Universitat 
Aut\'onoma de Barcelona, \\ Facultat de Ci\`encies, Torre C5, E-08193 Bellaterra 
(Barcelona), Spain\\
$^3$Departament d'Estructura i Constituents de la Mat\`eria,
Universitat de Barcelona, \\
Diagonal 647, 08028 Barcelona, Spain\\
$^4$Departamento de F{\'\i}sica At\'omica, Molecular y Nuclear, 
Universidad de Granada, E-18071 Granada, Spain \\
$^5$Instituto de F{\'\i}sica Corpuscular (centro mixto CSIC-UV),
Institutos de Investigaci\'on de Paterna, \\  Aptdo. 22085, 46071, Valencia, Spain}
}

\begin{abstract}
Charmed mesons in dense matter are studied within a unitary coupled-channel approach  which takes into account Pauli-blocking effects and meson self-energies in a self-consistent manner. We obtain the open-charm meson spectral functions in this dense medium, and discuss their implications on hidden charm and charm scalar resonances and on the formation of $D$-mesic nuclei.
\end{abstract}

\maketitle


\section{Introduction}
The properties of open and hidden charm mesons in a hot dense environment are being the focus of recent analysis. Indeed, part of the  physics program of the PANDA and CBM experiments at the  future FAIR facility
at GSI \cite{gsi00} will search, among others, for in-medium modification of hadrons in the charm sector,  providing a first insight into the charm-nucleus
interaction.  

The in-medium modification of the properties of open-charm mesons may lead to formation of $D$-mesic nuclei \cite{tsushima99}, and will also affect the renormalization of charm and hidden-charm scalar hadron resonances in nuclear matter, providing information not only about their nature but also about the interaction of the open-charm mesons with nuclei. In the present paper we obtain the  open-charm spectral functions in dense matter within a self-consistent approach in coupled channels, and analyze the effect on the properties of dynamically-generated charm and hidden charm scalar resonances and provide some insight into the formation of $D$-nucleus bound states.

\vspace{-0.4cm}

\section{Charmed mesons in matter}

The self-energy in symmetric nuclear matter for open-charm pseudoscalar ($D$) and vector ($D^*$) mesons is obtained following a self-consistent coupled-channel procedure. The transition potential of the Bethe-Salpeter equation for different isospin, total angular momentum, and finite density and temperature is derived from effective lagrangians, which will be discussed in the next subsection. The $D$ and $D^*$ self-energies are then obtained summing the transition amplitude for the different isospins over the nucleon Fermi distribution at a given temperature (see details in Refs.~\cite{TOL07,tolos09}). Then, the meson spectral function  reads
\begin{eqnarray}
&&S_{D(D^*)}(q_0,{\vec q}, \rho, T)= \nonumber \\
&& -\frac{1}{\pi}\frac{{\rm Im}\, \Pi_{D(D^*)}(q_0,\vec{q},\rho, T)}{\mid
q_0^2-\vec{q}\,^2-m_{D(D^*)}^2- \Pi_{D(D^*)}(q_0,\vec{q},\rho, T) \mid^2 } \ , \ \  \ \ \  \ \ 
\label{eq:spec}
\end{eqnarray}
with $\Pi_{D(D^*)}$ being the self-energy at given energy $q_0$, momentum $\vec{q}$, density $\rho$ and temperature $T$.

\vspace{-0.5cm}

\subsection{SU(4) and SU(8) schemes}
\label{su4}

\begin{figure}
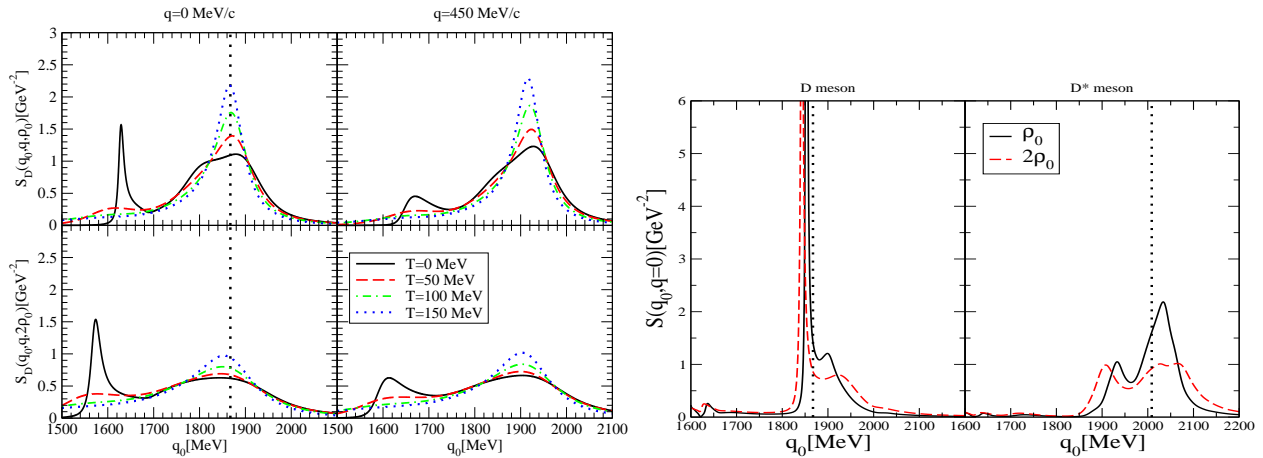

\includegraphics[width=0.5\textwidth, height=6cm]{paper_spectral_tot}
\hfill
\includegraphics[width=0.5\textwidth, height=5cm]{art_spec}
\caption{Left: $D$ meson spectral function for the SU(4) model. Right: $D$ and $D^*$ spectral functions in the SU(8) scheme. We show the $D$ and $D^*$ meson free masses for reference (dotted lines). \label{fig1}}
\end{figure}


The open-charm  meson spectral functions are obtained from the Bethe-Salpeter equation in coupled-channels taking, as bare interaction, two kinds of bare potential.

First, we consider a type of broken $SU(4)$ $s$-wave Weinberg-Tomozawa (WT) interaction supplemented by an attractive isoscalar-scalar term and using a cutoff regularization scheme. We fix this cutoff by generating dynamically the $I=0$ $\Lambda_c(2595)$ resonance. A new resonance in $I=1$ channel, $\Sigma_c(2800)$, is  generated \cite{LUT06,mizutani06}. 

The in-medium solution incorporates Pauli blocking, baryon mean-field bindings and meson self-energies \cite{TOL07}. In l.h.s. of Fig.~\ref{fig1} we display the $D$ meson spectral function for different momenta, temperatures and densities. At $T=0$ the spectral function shows two peaks: the $\Lambda_c(2595) N^{-1}$ and the quasi(D)-particle  peak  mixed with the $\Sigma_c(2880) N^{-1}$. Those states dilute with increasing temperature while the quasiparticle peak gets closer to its free value. Finite density results in a broadening of the spectral function because of the increased phase space.

Secondly, heavy-quark symmetry (HQS) is implemented by treating on equal footing heavy pseudoscalar and vector mesons, such as the $D$ and $D^*$ mesons. The $SU(8)$ WT includes pseudoscalars and vector mesons together with $J=1/2^+$ and $J=3/2^+$ baryons \cite{magas09,gamermann10}.  This symmetry is, however, strongly broken in nature and we adopt the physical hadron masses and different weak non-charmed and charmed pseudoscalar and vector meson decay constants. We also improve on the regularization scheme in matter beyond the cutoff method \cite{tolos09}.

In this scheme, all resonances in the $SU(4)$ model are reproduced and new resonant states are generated \cite{magas09} due to the enlarged Fock space. However, the nature of some of those resonances is different regarding the model. While the $\Lambda_c(2595)$ emerges as a $DN$ quasibound state in the $SU(4)$ model, it becomes predominantly a $D^*N$ quasibound state in the $SU(8)$ scheme.

The modifications of these resonances in the nuclear medium strongly depend on the coupling to $D$, $D^*$ and $N$ and are reflected in the spectral functions. On the r.h.s of  Fig.~\ref{fig1} we display the $D$ and $D^*$ spectral functions, which show then a rich spectrum of resonance ($Y_c$)-hole($N^{-1}$) states.  As density increases, these $Y_cN^{-1}$ modes tend to smear out and the spectral functions broaden with increasing phase space, as seen for the $SU(4)$ model \cite{mizutani06}.

\begin{figure}
\begin{minipage}{0.33\textwidth}
\includegraphics[height=5.5cm]{ds02317}
\end{minipage}
\hfill
\begin{minipage}{0.33\textwidth}
\includegraphics[height=5.5cm]{x37}
\end{minipage}
\hfill
\begin{minipage}{0.33\textwidth}
\includegraphics[height=6.3cm, angle=-90]{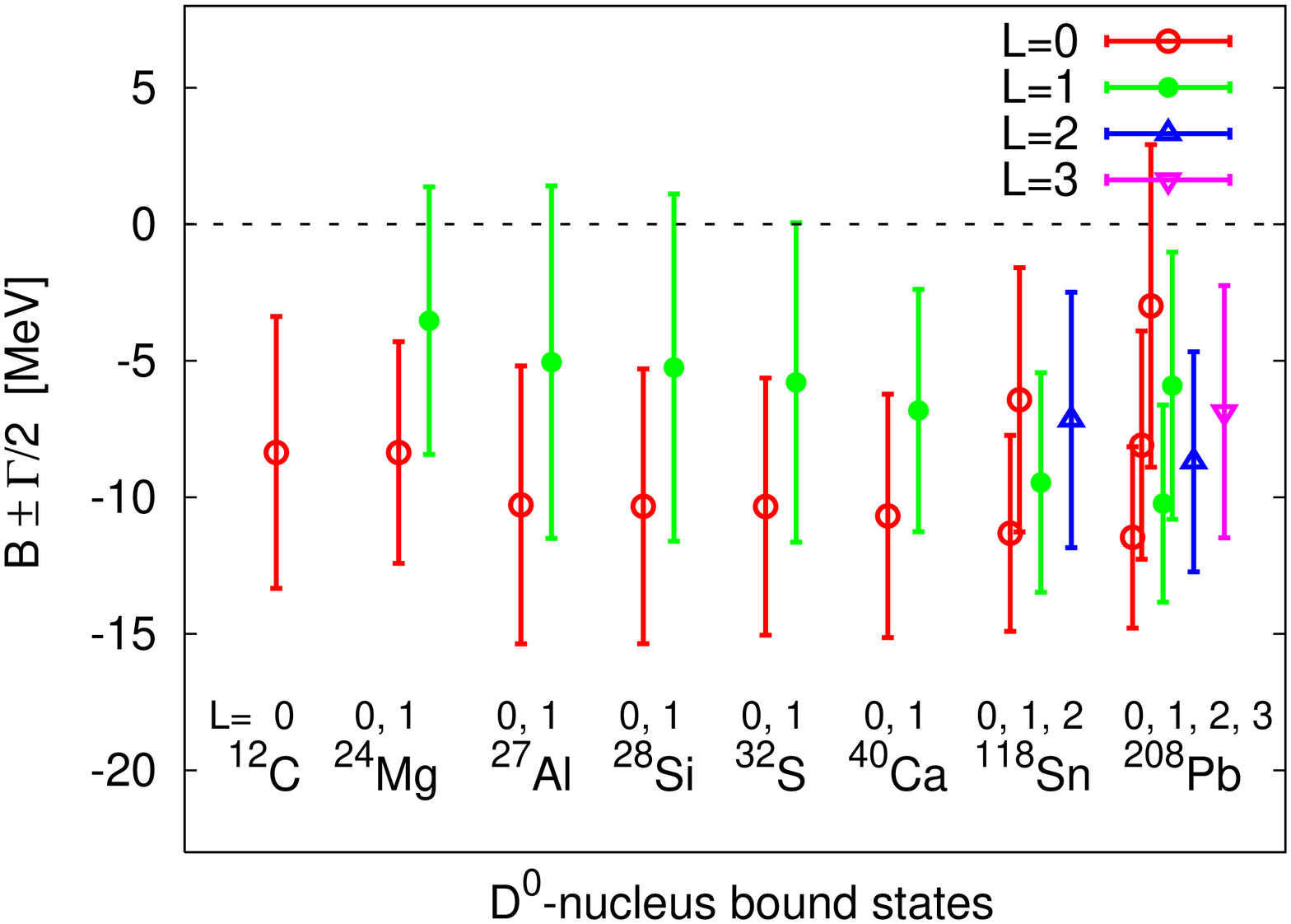}
\end{minipage}
\caption{$D_{s0}(2317)$ (left) and  $X(3700)$ (middle) resonances in nuclear matter. $D^0$ nucleus bound states (right) \label{fig2}}
\end{figure}


\vspace{-0.4cm}

\section{Scalar resonance in  matter}

The analysis of the properties of scalar resonances in nuclear matter is crucial in order to understand their nature, whether they are $q \bar q$, molecules, mixtures of $q \bar q$ with meson-meson components, or dynamically generated resonances from meson-meson scattering.

In the following we study the charmed resonance $D_{s0}(2317)$ \cite{Kolomeitsev:2003ac,guo06,Gamermann:2006nm} together with a hidden charm scalar meson, $X(3700)$, predicted in Ref.~\cite{Gamermann:2006nm}, which might have been observed  \cite{Gamermann:2007mu} by the Belle collaboration \cite{Abe:2007sy}. Those resonances are generated dynamically by solving the coupled-channel Bethe-Salpeter equation for two pseudoscalar mesons \cite{Molina:2008nh}. The $D_{s0}(2317)$ mainly couples to the $DK$, while the hidden charm state $X(3700)$ couples most strongly to $D\bar{D}$. Thus, any change in the $D$ meson properties in nuclear matter will have an important effect on these  resonances. The $D$ meson self-energy is given in the $SU(4)$ model without the phenomenological isoscalar-scalar term, but supplemented by the $p$-wave self-energy \cite{Molina:2008nh}.

 In Fig.~\ref{fig2} the $D_{s0}(2317)$ (left) and $X(3700)$ (middle) resonances are displayed via the squared transition amplitude for the corresponding dominant channel at different densities. The $D_{s0}(2317)$ and $X(3700)$ resonances, which have a zero and small width,
develop widths of the order of 100 and 200 MeV at normal nuclear matter density,  respectively. This is due to the opening of new many-body decay channels as the $D$ meson gets absorbed in the nuclear medium via $DN$ and $DNN$ inelastic reactions. We do not extract any clear conclusion for the mass shift. We suggest to look at the transparency ratio to investigate those in-medium widths, since it is very sensitive to the in-medium width of the resonance.

\vspace{-0.4cm}

\section{D-mesic nuclei}

$D$-meson bound states in $^{208}$Pb were predicted \cite{tsushima99} relying upon an attractive  $D$ meson potential. The observation of those bound states might be, though, problematic due to their widths, as in the case of the SU(4) model \cite{TOL07}. However, for the scheme with HQS \cite{tolos09} the $D$ meson in nuclear matter has a sufficiently small width with respect to the mass shift to form bound states in nuclei.

In order to compute de $D$-nucleus bound states, we solve the Schr\"odinger equation. We concentrate on $D^0$-nucleus bound states \cite{carmen10}. The potential that enters in the equation is an energy-dependent one that results from the zero-momentum $D$-meson self-energy within the SU(8) model \cite{tolos09}. In Fig.~\ref{fig2} (right) we show  $D^0$ meson bound states in different nuclei. We observe that the $D^0$-nucleus states are weakly bound, in contrast to previous results \cite{tsushima99}.  Their experimental detection is, though, difficult.

\vspace{-0.4cm}

\section{Conclusions}

Open-charm mesons ($D$ and $D^*$) in dense matter have been studied within a self-consistent coupled-channel approach taking, as bare interaction, different effective lagrangians. The in-medium solution  accounts for Pauli blocking effects and meson self-energies. We have analyzed the evolution in matter of the open-charm meson spectral functions and discussed their effects on the  $D_{s0}(2317)$ and the predicted $X(3700)$ in nuclear matter, and suggested to look at transparency ratios to investigate the in-medium width of those resonances. We have finally analyzed the possible formation of $D$-mesic nuclei. Only  weakly bound $D^0$-nucleus states seem to be feasible within the SU(8) scheme that incorporates heavy-quark symmetry. However, its experimental detection is most likely a challenging task.

\vspace{1cm}

L.T. acknowledges support from the RFF program of the University of Groningen. This work is partly supported by the EU contract No. MRTN-CT-2006-035482 (FLAVIAnet), by the contracts FIS2008-01661 and FIS2008-01143 from MICINN (Spain), by the Spanish Consolider-Ingenio 2010 Programme CPAN (CSD2007-00042), by the Generalitat de Catalunya contract 2009SGR-1289 and by Junta de Andaluc\'{\i}a under contract FQM225. We acknowledge the support of the European Community-Research Infrastructure Integrating Activity ``Study of Strongly Interacting Matter'' (HadronPhysics2, Grant Agreement n. 227431) under the 7th Framework Programme of EU.


\vspace{-0.4cm}

\bibliographystyle{aipproc}

\end{document}